# Magnetic Texture in Insulating Single Crystal High Entropy Oxide Spinel Films

*Yogesh Sharma[†,#,\*], Alessandro R. Mazza[†], Brianna L. Musico[§], Elizabeth Skoropata[†], Roshan Nepal[¶], Rongying Jin[¶], Anton V. Ievlev[‡], Liam Collins[‡], Zheng Gai[‡], Aiping Chen[#], Matthew Brahlek[†], Veerle Keppens[§], Thomas Z. Ward[†,\*]*

[†]*Materials Science and Technology Division, Oak Ridge National Laboratory, Oak Ridge, TN 37831, USA*

[#]*Center for Integrated Nanotechnologies (CINT), Los Alamos National Laboratory, Los Alamos, NM 87545, USA*

[§]*Department of Material Science and Engineering, University of Tennessee, Knoxville, TN 37996, USA*

[¶]*Department of Physics & Astronomy, Louisiana State University, Baton Rouge, LA 70803, USA*

[‡]*Center for Nanophase Materials Sciences, Oak Ridge National Laboratory, Oak Ridge, TN 37831, USA*

[\*]E-mail: ysharma@lanl.gov and wardtz@ornl.gov

**ABSTRACT**

Magnetic insulators are important materials for a range of next generation memory and spintronic applications. Structural constraints in this class of devices generally require a clean heterointerface that allows effective magnetic coupling between the insulating layer and the conducting layer. However, there are relatively few examples of magnetic insulators which can be synthesized with surface qualities that would allow these smooth interfaces and precisely tuned interfacial magnetic exchange coupling which might be applicable at room temperature. In this work, we demonstrate an example of how the configurational complexity in the magnetic insulator layer can be used to realize these properties. The entropy-assisted synthesis is used to create single crystal $(Mg_{0.2}Ni_{0.2}Fe_{0.2}Co_{0.2}Cu_{0.2})Fe_2O_4$ films on substrates spanning a range of strain states. These films show smooth surfaces, high resistivity, and strong magnetic responses at room temperature. Local and global magnetic measurements further demonstrate how strain can be used to manipulate magnetic texture and anisotropy. These findings provide insight into how precise magnetic responses can be designed using compositionally complex materials that may find application in next generation magnetic devices.

## 1. INTRODUCTION

Magnetic insulators (MI) are promising materials for spintronics, spincaloritronics, nonvolatile memories, and microwave applications.[1–10] The selection of the MI heterostructures in a single crystal thin film, grown on a suitable substrate, are critical for realizing quality interfaces for these applications.[2,11] However, excellent quality MI single crystalline materials of a Curie temperature well above room temperature are rare. Ferrimagnetic insulating oxides with the $AB_2O_4$ spinel structure, have been widely studied for these applications as they offer interesting MI properties.[1,12–17] Within spinel oxides there are three different structural motifs: normal, random and inverse, which are distinguished based on the distribution of cations among the tetrahedral and octahedral sites.[12,18–20] In particular, inverse spinel ferrites (B = Fe) have received a great deal of interest as room temperature MI for the aforementioned applications.[1–8] The entropy change associated with the cation disordering in a spinel configuration has a major impact on the thermodynamic landscape required to stabilize these spinel structures.[21,22] Therefore, the cation disorder at tetrahedral and octahedral sites is a critical factor in determining magnetic, optical, and transport properties.[12,19,23] The dependence of the configurational entropy on the degree of cation disorder has been reported in spinel ferrites as leading to variation in magnetic ground states.[24–27] To take advantage of this complex cation disorder, chemical doping is widely used to tune magnetic functionality in spinel ferrites.[28–30] However, previous studies show that incorporating multiple chemical dopants into the *A*- or *B*-site in spinel ferrites can lead to quenched chemical mixing, which drives the creation of undesired secondary phases and poor surface morphologies which preclude epitaxial heterostructuring.[31–33] Given the complexity of the single phase formation in multi-cation doped-spinel ferrites where a random cation distribution can be used to tailor magnetic properties, synthesizing spinels which can be made to host a large number of different cations on the *A*- and/or *B*-sites without disrupting crystallinity or structural phase would be an exceptional tool towards designing MIs.[4,34]

The development of entropy-assisted synthesis in oxide systems is a promising approach to manipulate the cation combinations hosted in the spinel ferrite lattice.[35–37] Unlike high-entropy alloys, where metal ions are held in a metal-metal bonded solid solution, the high entropy oxides are comprised of ionic and covalently bonded cations and anions.[38,39] This effectively maximizes the number of microstates and thereby helps to drive homogeneous distribution of cations through the cation sublattice(s).[38,40–45] For magnetism, this approach can be especially beneficial as the localization of electrons and more complex exchange interactions in this bonding environment can be integrated into magnetic design strategies.[39,46–49]

Here, we use entropy-assisted synthesis to create high entropy, single-crystal configurationally complex spinel oxide (CCSO) films of the $AB_2O_4$ type spinel ferrite $(Mg_{0.2}Ni_{0.2}Fe_{0.2}Co_{0.2}Cu_{0.2})Fe_2O_4$ in different strain states. At room temperature, these films are found to be highly insulating and magnetic with smooth surfaces. Heteroepitaxial strain is found to have an important role in determining how magnetic texture and anisotropy develop. Room temperature magnetic force microscopy (MFM) demonstrates that strain engineering can be used to manipulate long-range magnetic stripe domain formation in CCSO films.

## 2. METHODS

**2.1 Epitaxial film growth.** A ceramic stoichiometric target of $(Mg_{0.2}Ni_{0.2}Fe_{0.2}Co_{0.2}Cu_{0.2})Fe_2O_4$ (CCSO) was synthesized using the conventional solid-state reaction method. Pulsed laser epitaxy was used to grow CCSO thin films on the three different substrates, including MgO ($a$ = 4.212 Å), spinel $MgAl_2O_4$ (MAO, $a/2$ = 4.040 Å), and $SrTiO_3$ (STO, $a$ = 3.905 Å). Before loading the substrate into growth chamber, the substrate cleaning was performed using following steps; (1) substrates are cleaned with acetone and methanol in an ultrasonic bath for 2 min each to remove dust particles and (2) the substrates are soaked in deionized (DI) water for 4-5 min in an ultrasonic bath following by blow drying the substrate with nitrogen gas. A KrF excimer laser ($\lambda$ = 248 nm) operating at 10 Hz was used for target ablation. The laser fluence was 1.2 J/cm$^2$ with an area of 3.5 mm$^2$ on the target. The target-substrate distance was set at 5 cm. Deposition optimization was performed and the optimal growth conditions were found to occur with an oxygen partial pressure of 30 mTorr at a substrate temperature of 670 °C. Thin film samples of each strain state were grown. After deposition, the films were cooled to room temperature under 100 Torr oxygen pressure.

**2.2 Measurements.** The crystal structure and growth orientation of the films were characterized by X-ray diffraction (XRD) using a four-circle high resolution X-ray diffractometer (X'Pert Pro, Panalytical) (Cu K$\alpha_1$ radiation). We used x-ray reflectivity (XRR) measurements to measure the thickness of the films. Magnetization as a function of temperature and applied field, were recorded using a Quantum Design MPMS3 superconducting quantum interference device (SQUID) magnetometer. Isothermal magnetization curves were obtained for magnetic fields: $-7\ T \leq H \leq 7\ T$ at T= 300 K. X-ray absorption and magnetic circular dichroism were measured at the Advanced Photon Source at Argonne National Laboratory at beamlines 4-ID-C. MFM measurements were performed on an Asylum Research Cypher microscope using Co/Cr-coated (Asylum Research, ASYMFMHM) atomic force microscope (AFM) tips, having a resonance frequency and spring constant of 70 kHz and 2.0 N/m, respectively. The tips were magnetized, prior to start scanning the samples, by bringing them into close proximity with a strong magnet. A photo-thermal excitation (i.e. Blue Drive Excitation) mechanism was used to achieve a clean mechanical excitation of the cantilever. The MFM images were recorded in dual pass mode at 100 nm lift height using phase sensitive detection. ToF-SIMS measurements were performed using TOF.SIMS.5-NSC (ION.TOF GmbH, Germany) instrument. We used a Bi$_3^+$ liquid metal ion gun (for imaging mode; energy 30 keV, current 0.5 nA, spot size ~120 nm, and for spectral mode; energy 30 keV, current 30 nA, spot size ~5 μm) as a primary ion source. A Cs$^+$ ion gun was used as a sputter source (energy 1 keV, current 70 nA, spot size ~20 μm). Secondary ions were analyzed using time-of-flight mass analyzer operated in positive ion detection mode with mass resolution m/Δm= 3000–7000 (for spectral mode) and 100–300 (for spectral mode) to track peaks of Fe$^+$, Mg$^+$, Co$^+$, Cu$^+$, O$^+$, Ni$^+$, and MgO$^+$. Surface measurements were carried out using Bi$^{3+}$source, rastered over 25x25 μm$^2$ area with 256x256 px resolution. 3D measurements were performed in non-interlaced mode, when each scan by Bi$^{3+}$ primary source was followed by sputtering using Cs$^+$ source for 2s (over 100x100 μm area). Acquired three-dimensional maps of the peaks' spatial distribution were used to identify local changes in the chemistry of the studied sample.

## 3. RESULTS AND DISCUSSION

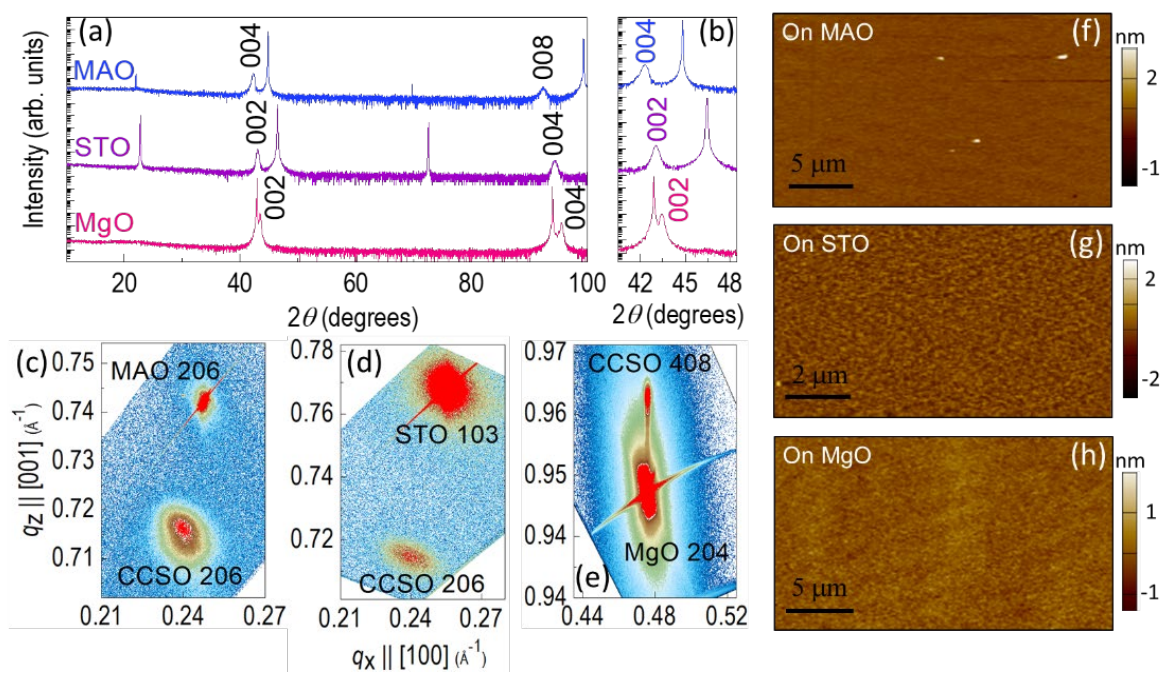

**Figure 1.** Epitaxial growth of configurationally complex spinel oxide (CCSO) films. (a) X-ray diffraction $\theta$-$2\theta$ scans of (Mg$_{0.2}$Ni$_{0.2}$Fe$_{0.2}$Co$_{0.2}$Cu$_{0.2}$)Fe$_2$O$_4$ films grown on MAO, STO, and MgO substrates with the film thicknesses of 68, 70 and 64 nm, respectively. In all CCSO films, only the substrate and film peaks corresponding to their (00$l$) planes appear. (b) The $\theta$-$2\theta$ scans around the substrate 002 peaks (for STO and MgO) and 004 (for spinel MAO), where the Bragg's peaks corresponding to films are denoted. Reciprocal space maps of asymmetric scans around the (c) MAO 206, (d) STO 103 and (e) MgO 204 peaks show relationship between films

and substrates. (f-h) Atomic force microscopy images shows smooth surfaces, especially for the strained films grown on MAO and MgO substrates.

Figures 1a and 1b show X-ray diffraction (XRD) $\theta$-$2\theta$ scans for CCSO films deposited on MgO, SrTiO$_3$ (STO) and MgAl$_2$O$_4$ (MAO) substrates. Only 00$l$ reflections of the CCSO films are observed in the $\theta$-$2\theta$ scans, indicating that all films are epitaxial and phase pure. The in-plane epitaxial orientation relationship of the films to the substrates is cube-on-cube to the (001)-oriented MgO, STO, and MAO substrates: (001) CCSO ∥ (001) MgO/STO/MAO; [100] ∥ [100], and the rocking curves on the 002/004 peaks in $\omega$ show full width at half maximum (FWHM) of 0.07°, 0.8°, 0.6° for films on MgO, STO and MAO substrates, respectively (Figure S1, supporting information). Figures 1c–e show reciprocal space maps (RSMs) around the 206 (for MAO), 103 (for STO) and 204 (for MgO) peaks and the CCSO films' peaks. Using the bulk lattice parameter found in the bulk ceramic CCSO target of around 8.36 Å,[35] the applied lattice mismatches between the film and the substrate are given as -3.35%, -6.58% and +0.76%, for MAO, STO and MgO substrates, respectively. The films grown on MgO is coherently tensile-strained. The film grown on MAO is compressively strained, showing some relaxation away from the substrate at the 68 nm thicknesses presented here, whereas the film on STO is completely relaxed. Figures 1f–h show the surface morphologies of these films captured using atomic force microscopy (AFM). The AFM images show very smooth surfaces with the root mean square surface roughness of 0.28, 0.98 and 0.35 nm on MgO, STO and MAO, respectively.

In Figures 2a and 2b, we show room-temperature Raman spectra collected on CCSO films grown on MAO and MgO substrates. Data for the film grown on STO is not shown as the substrate's Raman background is too strong to separate the CCSO film's contributions. The Raman spectra from the substrates (MgO and MAO) and epitaxial CoFe$_2$O$_4$ (CFO) films are also presented in Figures 2a and 2b to provide a reference to a less complex magnetic insulating spinel. CFO shows an inverse spinel structure in which all Co-ions occupy the octahedral sites and the Fe-ions are distributed almost equally between the tetrahedral and octahedral sites. The Raman spectrum of the CCSO film on MgO shows six phonon modes at 328, 471, 556, 612, 662, and 702 cm$^{-1}$ (Figure 2b), which are consistent with the phonon modes observed in an inverse spinel CFO film and MgFe$_2$O$_4$ bulk single crystal.[16,50]

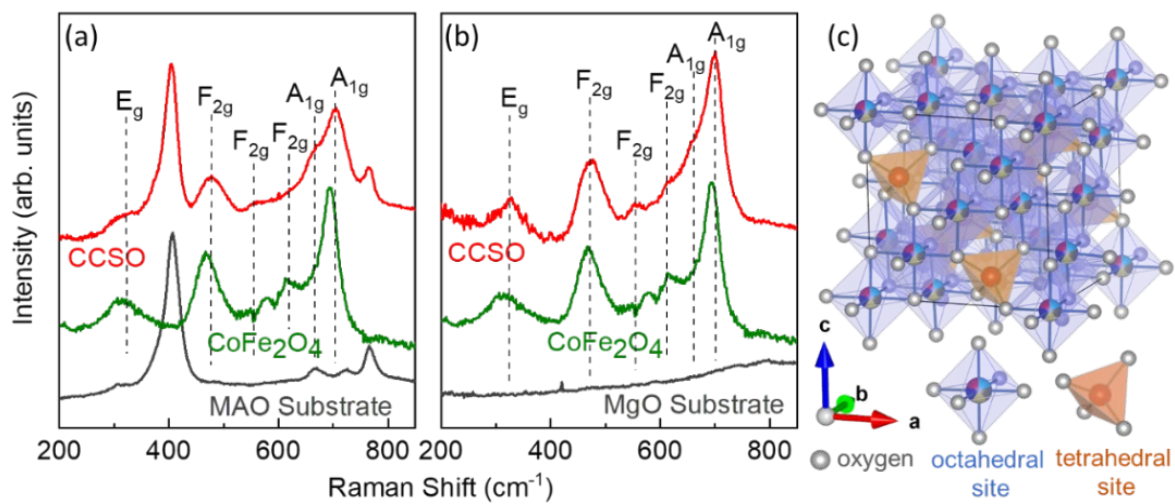

**Figure 2.** Room temperature Raman spectra of CCSO films. The Raman spectra of CCSO films grown on (a) MAO and (b) MgO substrates, including the spectra from each substrate and CoFe$_2$O$_4$ epitaxial film for reference. Raman spectroscopy measurements show the presence of all phonon modes akin to CoFe$_2$O$_4$-inverse spinel ferrite structure. (c) Schematic of inverse spinel CCSO's crystal structure with octahedral and tetrahedral sites.

The inverse spinel structure of CCSO films is confirmed by representative phonon modes observed by micro-Raman spectroscopy measurements while comparing with the phonon modes from the CFO film, which can follow the expected ordering of $(Fe_1^{3+})_{Tetra}[Fe_1^{3+}Ni_{0.2}^{2+}Cu_{0.2}^{2+}Fe_{0.2}^{2+}Mg_{0.2}^{2+}Co_{0.2}^{2+}]_{Octa}O_4$, based on site-preference energies for the constituent cations.[21,51,52] Further, x-ray absorption spectroscopy and x-ray magnetic circular dichroism measurements of Fe, Co, Ni, and Cu *L*-edges, shown in Figure S2 for strained CCSO films on MAO and MgO, indicate that the Fe ions are located at both octahedral (Fe$^{3+}$ and Fe$^{2+}$) and tetrahedral (Fe$^{3+}$) sites whereas other transition metal ions appear to be all at octahedral site. Notice that the phonon modes corresponding to CCSO films are at higher wavenumber (blue-shifted) than those of the CFO modes. This is consistent with the large difference in the ionic radii of cations present at octahedral and tetrahedral sites.[50] Interestingly, at higher wavenumber (≥ 400 cm$^{-1}$) in Raman spectra, the blue-shift in phonon modes is more pronounced for the CCSO film on MAO compared to the film on MgO, which might be due to the different lattice-strain and tetragonality distortion in the film[16,53], as evident by the tetragonality ratio and lattice parameters obtained from XRD-RSM data (Figure S1c). Raman results indicate the strain modification of lattice vibrations in inverse spinel CCSO films, which can be reflected in exchange coupling pathways and the functionalities sensitive to the interatomic bond angles and bond lengths, such as magnetic and dielectric responses.[16,53]

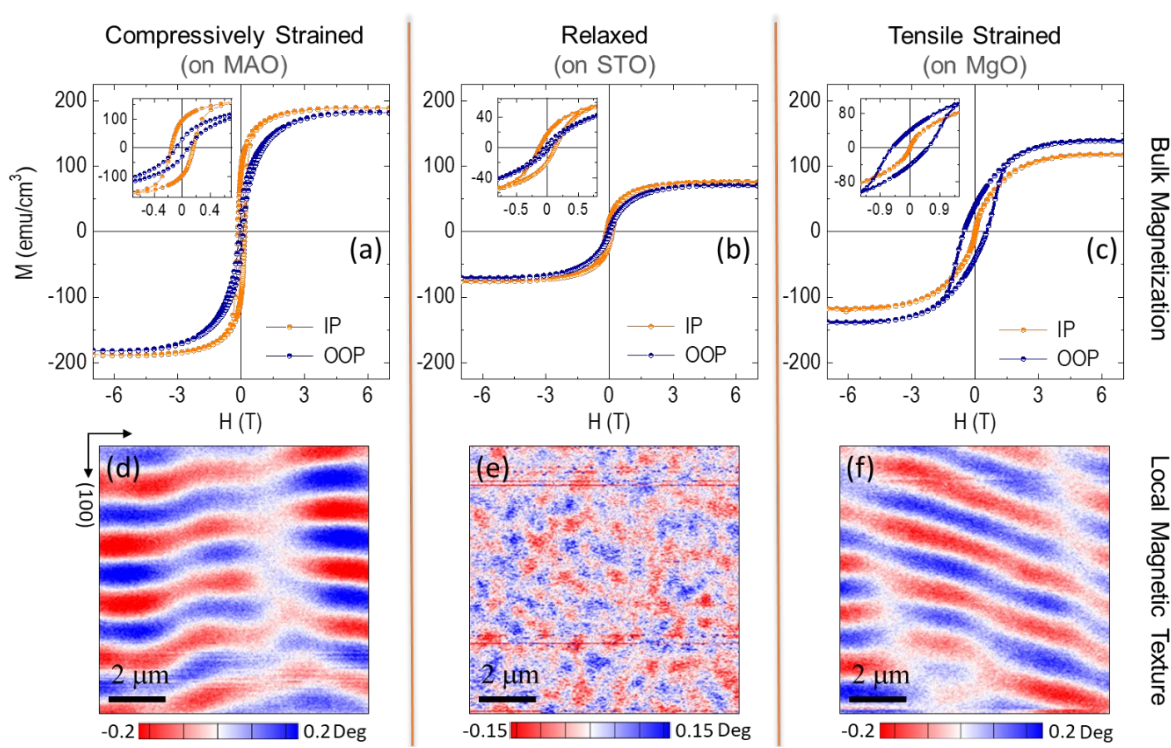

**Figure 3.** Room temperature magnetism and the role of lattice-strain on magnetic behavior of CCSO films. (a-c) Room temperature magnetization loops with magnetic field directed in-plane (IP) and out-of-plane (OOP) to the sample's c-axis. The graphs are plotted on the same vertical scale to emphasize the influence of the strain imparted by the different substrates. The insets are the zoomed-in view of the magnetization loops. MFM images recorded for CCSO films grown on (d) MAO, (e) STO and (f) MgO at room temperature and at zero applied magnetic field. The scan size for each image is 10μm×10μm. Strained CCSO films on MgO and MAO substrates show a stripe domain pattern, whereas the relaxed CCSO film on STO shows magnetic puddle-like domains.

Figures 3a–c show room temperature magnetic hysteresis loops for CCSO epitaxial films. The temperature-dependent magnetization measurements performed on the CCSO films show a magnetic transition onset above room temperature regardless of strain state (Figure S3a–c, supporting information). Strain effects appear to have a strong influence on absolute magnetization, with both the compressive and tensile strain pushing the magnetization to higher saturation moment. For the compressively strained CCSO film on the spinel MAO substrate (Figure 3a), we observe a higher magnetization compared to the CCSO films on STO (Figure 3b) and MgO (Figure 3c) substrates. For the tensile-strained CCSO film grown on MgO (Figure 3c), an out-of-plane hysteresis loop with a coercive field of 5.5 kOe is observed, while the coercivity along the in-plane direction is found to be 0.06 kOe (inset of Figure 3c and Table S1, supporting information). This indicates that the easy axis is orientated along the out-of-plane direction for the CCSO film grown on MgO. In the case of the partly relaxed and fully relaxed films on MAO and STO substrates, respectively, a different magnetic behavior is observed; the out-of-plane hysteresis loop is greatly reduced and the magnetic easy axis moves toward the in-plane direction. For the relaxed CCSO film on STO, coercive fields of 0.4 and 1.3 kOe are observed along the out-of-plane and in-plane directions, respectively (inset of Figure 3b and Table S1, supporting information). The magnetization results show that the lattice anisotropy can be used to modify magnetic behavior in CCSO films. The magnetic easy axis is along the in-plane direction for the relaxed and compressively strained films whereas, tensile strain results in a magnetic easy axis along the out-of-plane direction. Crystal symmetry changes likely cause the observed differences in saturation moment between strained and unstrained films. As there is no obvious change in charge state of the cations, a possible explanation is that strain drives crystal field splitting that pushes some transition metal sites to a higher spin state (Figure S1, supporting information). Follow on magnetic and structural correlation studies such as those accessible with polarized neutron reflectometry will be required to address the role of surface and shape anisotropy in driving the emergence of the observed magnetic texture. Moreover, in all cases, the films are highly insulating at room temperature and the resistivity values reached the instrument limit, close to 360 K. (Figure S3d, supporting information).

To begin to understand how these macroscopic magnetic properties are locally influenced by configurational disorder and lattice-strain, MFM is performed at room temperature to visualize the emergence of the local magnetic structure (domains and domain walls). Figures 3d–f show that the application of both tensile and compressive strains lead to the formation of micron-scale stripe domains, where the periodicity and variation of the phase change represented by the color scale indicate domains and magnetic orientations. The average width of the domains is measured to be around 1 μm. The alternated distributed blue and red stripes indicate that spins in neighboring domains are in an antiparallel arrangement. Magnetic puddle-like domains with random magnetic alignment between them are observed in the relaxed CCSO film (Figure 3e). Comparing topography imaged during MFM and the section analysis performed on the same regions of topography and MFM images show no correlation between the film surface and morphology to stripe domain pattern (Figure S4, supporting information). Moreover, the nature of magnetic stripe domains before and after heating above Curie temperature (800°C) and acquiring corresponding surface topography images indicate the formation of robust and hard magnetic stripe domains (Figure S5, supporting information).

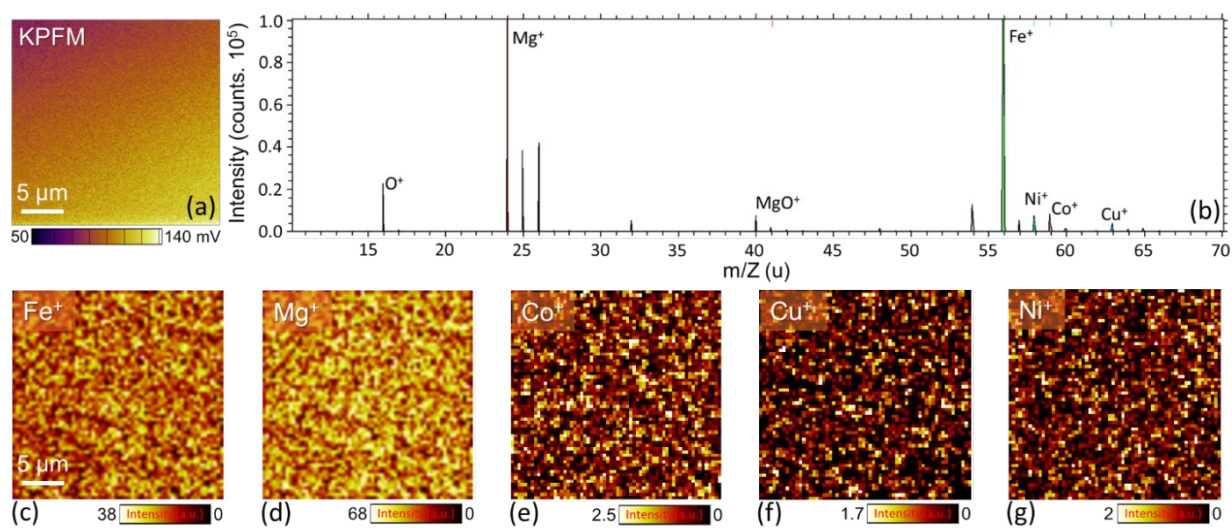

**Figure 4.** Chemical imaging of strained CCSO film. (a) KPFM image recorded on the surface of CCSO film (on MgO) shows no significant change in surface chemical potential profile. (b) A full averaged mass spectrum measured from ToF-SIMS confirms the presence of Mg, Fe, Ni, Co, Cu, and O ions in the CCSO film. Note that the extra peaks in spectrum are related to the extrinsic surface contamination unrelated to sample composition. (c-g) Chemical maps of spatial distribution of base elements Fe, Mg, Co, Cu, and Ni recorded in positive ion detection mode measured on the same sample by ToF-SIMS show excellent mixing and no elemental clustering.

To confirm that these magnetic stripe domains are not induced by chemical segregation on the top surfaces of the films, the local chemical profiles on the surfaces and bulk of the films are investigated by Kelvin probe force microscopy (KPFM) and time-of-flight secondary ion mass spectroscopy (ToF-SIMS). In Figure 4a, a KPFM image recorded over 25μm×25μm area of the surface of the CCSO film on MgO shows no appreciable change in the surface chemical potential profile. This indicates that the stripe domains observed in the MFM are not the result of any long-range chemical segregation. This is also confirmed in the bulk using ToF–SIMS to investigate the local chemical composition of the CCSO film grown on MgO. The chemical sensitivity of SIMS allows the detection of the ion distribution with a spatial resolution of ~120 nm in the plane[8], which would easily allow one to resolve any possible chemical stripe segregation on the length scale observed in the MFM if present. Figure 4b shows an averaged mass spectrum with all the constituent elements $Mg^+$, $Fe^+$, $Co^+$, $Ni^+$ and $Cu^+$ present in the film. In this ToF–SIMS setting, the distribution of the corresponding peak area as a function of spatial location permits characterization of local chemical changes within the studied area.[8,54] Figures 4c–g show element specific chemical mappings of $Fe^+$, $Mg^+$, $Co^+$, $Cu^+$, and $Ni^+$ ions, respectively. These chemical maps further confirm that there is no chemical segregation as a function of lateral position in the film. After acquiring these data, the film surface is sputtered using a rastered $Cs^+$ cluster beam of 1keV energy by creating a 2-3 nm deep and 100 μm×100 μm square crater onto the film. The chemical maps across this crater also do not show any sign of local chemical segregation, confirming the chemical homogeneity of the film even below the surface (Figure S6, supporting information).

## 4. CONCLUSIONS

In summary, we have synthesized single crystal heteroepitaxial films of the configurationally complex spinel oxide $(Mg_{0.2}Ni_{0.2}Fe_{0.2}Co_{0.2}Cu_{0.2})Fe_2O_4$. These materials show themselves to be room temperature magnetic insulators that are highly sensitive to strain. Long range magnetic stripe domains are observed in strained films at room temperature and the domain configurations are found to be highly dependent on the lattice symmetry modification induced by the underlying substrate. Configurationally complex spinels open a pathway to access exceptional tunability over spin and magnetic exchange interactions that are inaccessible in less complex spinels hosting only one or two cations.

## ASSOCIATED CONTENT

**Supporting information:** X-ray diffraction rocking curves and phi-scan results, X-ray absorption spectroscopy (XAS) and x-ray magnetic circular dichroism (XMCD) measurements, temperature dependent magnetization and resistivity measurements, surface topography together with MFM images and the MFM measurements after and before heating the sample above Curie temperature.

## AUTHOR INFORMATION

**Corresponding author**

**Yogesh Sharma** (ysharma@lanl.gov)

**T. Zac Ward** (wardtz@ornl.gov)

**Notes: The authors declare no competing financial interest.**

## ACKNOWLEDGMENTS

Experimental design, synthesis, structural and magnetic characterizations were supported by the U.S. Department of Energy (DOE), Office of Science, Basic Energy Sciences (BES), Materials Sciences and Engineering Division. Scanning probe microscopy and TOF-SIMS were performed as user projects at the Center for Nanophase Materials Sciences, which is sponsored at Oak Ridge National Laboratory by the Scientific User Facilities Division, BES, U.S. DOE.

Y. S. acknowledges the support from the G. T. Seaborg Fellowship (project number 20210527CR) and the Center for Integrated Nanotechnologies, an Office of Science User Facility operated for the U.S. Department of Energy Office of Science at Los Alamos National Laboratory. R.N. and R.J. acknowledge the financial support by the U.S. National Science Foundation under Grant No. DMR-1504226. B.L.M. acknowledges the support of the Center for Materials Processing, a Tennessee Higher Education Commission (THEC) supported Accomplished Center of Excellence. This manuscript has been authored by UT-Battelle, LLC under Contract No. DE-AC05-00OR22725 with the U.S. Department of Energy. The United States Government retains and the publisher, by accepting the article for publication, acknowledges that the United States Government retains a non-exclusive, paid-up, irrevocable, world-wide license to publish or reproduce the published form of this manuscript, or allow others to do so, for United States Government purposes. The Department of Energy will providepublic access to these results of federally sponsored research in accordance with the DOE Public Access Plan (http://energy.gov/downloads/doe-public-access-plan).

**AUTHORS CONTRIBUTIONS**

Y.S. and T.Z.W. conceived and designed the experiments. Y.S. and B.L.M. fabricated the samples. Y.S., A.R.M., E.S. and Z. G. performed bulk magnetic measurements. Y.S., R.N., R.J., and L.C. performed the MFM measurements. A.V.I. performed the ToF-SIMS measurements. The manuscript was written through contributions of all the authors including A.C., M.B., and V.K. All authors have read and agreed to the final version of the manuscript.